\documentstyle[aps,preprint,eqsecnum,epsf]{revtex}

\begin{document}
\draft

\title{
Nonclassical fields and the nonlinear interferometer
}

\author{ Barry C.\ Sanders and Dien A.\ Rice}
\address{
	Department of Physics, 
	Macquarie University, Sydney, New South Wales 2109, Australia
}
\date{\today}
\maketitle

$ $

\begin{abstract}

We demonstrate several new results for the nonlinear interferometer,
which emerge from a formalism which elegantly describes
the output field of the nonlinear interferometer as two-mode 
entangled coherent states.
We clarify the relationship between squeezing and 
entangled coherent states, since a weak nonlinear evolution 
produces a squeezed output, 
while a strong nonlinear evolution produces a two-mode, two-state
entangled coherent state.
In between these two extremes exist superpositions of two-mode
coherent states manifesting varying degrees of
entanglement for arbitrary values of the nonlinearity.  
The cardinality of the basis set of the entangled coherent states 
is finite when the ratio $\chi / \pi$ is rational,
where $\chi$ is the nonlinear strength.
We also show that entangled coherent states can 
be produced from product coherent states via a nonlinear medium
{\em without} the need for the interferometric configuration.
This provides an important experimental simplification in the process
of creating entangled coherent states. 

\end{abstract}

$ $


\newpage


\section{Introduction}

The nonlinear interferometer exhibits
remarkable properties, even for semiclassical input fields such as
a product coherent state, $|\alpha \rangle_a |\beta\rangle_b$ 
into the two input ports.  Despite the classical nature of
the input state, the output state can be highly nonclassical.  
This nonclassical nature is most apparent in the manifestation of the 
entangled coherent states.  We develop and apply an entangled
coherent states formalism to obtain new results.  Firstly, we
establish the relationship between entangled coherent states
and squeezed states, arising from weak nonlinearity, $\chi$,
where $\chi$ is proportional to the nonlinear parameter of the
medium and to the interaction time in the medium.  Strong
nonlinearity, in contrast, produces a two-mode, bi-valued 
entangled coherent state,
\begin{equation}
	2^{-1/2} (|0 \rangle_{a'} |\alpha \rangle_{b'}
	+ i |-i \alpha \rangle_{a'} |0 \rangle_{b'} ) .
\end{equation}
In between these two extreme nonlinearities, the output
is a general entangled coherent state, which interpolates
between the two-mode bi-valued entangled coherent state
and the squeezed state.  Moreover, for $\chi / \pi$ rational,
the output state is a finite sum of product coherent states,
and for $\chi / \pi$ irrational, the sum is replaced
by an integral.  Finally, entangled coherent states can
be produced {\em without} the need for the interferometer
configuration, resulting in a significant simplification
for producing entangled coherent states.

The construction of the nonlinear interferometer is achieved 
by placing a nonlinear optical medium along one internal optical 
path.  For example, a nonlinear medium can be placed in one or both arms 
of the Mach-Zehnder interferometer \cite{ref1,ref3,ref2,ref2b}.  
The ideal nonlinear Mach-Zehnder interferometer is mathematically 
equivalent to other nonlinear two-mode interferometers.
Loss and phase diffusion are assumed to be negligible in
this treatment.

We treat the nonlinear medium as a classical
object which enables photon-photon interactions.
The input field at each port is treated as a single-mode field.  In
Section II, we develop a formal treatment of the interferometer as a unitary
transformation of two input states into two output states.
In Section III, we discuss the nature of the interferometer 
and examine special cases and also the case for an arbitrary
value of the nonlinearity coefficient $\chi$.  
In Section IV, we derive the result that a weak value of
the nonlinearity coefficient produces squeezed state
outputs.  In Section V, we show an alternative
approach to producing entangled coherent states without using a nonlinear
interferometer.  In Section VI, we present our conclusions.

\section{Formalism}

The ideal nonlinear interferometer can be
described by a unitary transformation of the input fields into the
output fields.  There are two input fields $a$ and $b$ as shown in
Fig.\ 1, and the output fields are designated by $a'$ and $b'$.  The
input fields and output fields are treated as single-mode fields.

For simplicity we consider the special case that the two input fields at
the two input ports of the interferometer are coherent states \cite{ref9}
$ |\alpha\rangle = \hat{D}(\alpha)|0\rangle $
for
\begin{equation}
	\hat{D}(\alpha) = \exp ( \alpha \hat{a}^{\dag}
	 - \alpha^{*} \hat{a})
\label{displacement}
\end{equation}
the displacement operator.  The coherent field is the closest quantum
analogy to the classical coherent field.  Its properties include being
in a minimum uncertainty state and being generated by a classical
current distribution.  Coherent field inputs in each arm have proven to
be very interesting as these are precisely the inputs considered for
squeezing experiments \cite{ref4} and for obtaining entangled coherent
states 
\cite{ecs0,ecs1,ecs2,ecs3,ecs4,ecs5,ecs6,ecs7,ecs8},
also regarded as superpositions of multimode
coherent states \cite{ecs9,ecs10,ecs11,ecs12}.  
The unitary 
transformation operator for the nonlinear
interferometer is designated by $\hat{\cal I}$.  The
transformation is a sequence of a beamsplitter transformation $\hat{\cal B}$
followed by a path difference operator $\hat{\Delta}$ followed by the
commuting Kerr transformations in each arm, $\hat{\cal S}_1$ and $\hat{\cal S}_2$,
for which $[ \hat{\cal S}_1, \hat{\cal S}_2 ]=0$, and then a final beamsplitter
transformation $\hat{\cal B}$.  The net transformation is thus
\begin{equation}
	\hat{\cal I} = \hat{\cal B} \hat{\cal S}_1 \hat{\cal S}_2 \hat{\Delta} \hat{\cal B}
\end{equation}
where $\hat{\cal B}$, $\hat{\cal S}_1$ and $\hat{\Delta}$ are discussed below.

The 50/50 beamsplitter transformation is given 
by \cite{bs1,bs2,bs3,bs4,bs5,bs6,bs7}
\begin{equation}
	\hat{\cal B} = \exp \left( i \pi \left[ \hat{a}^{\dag} b
	+ \hat{a} b^{\dag} \right] /4 \right) .
\label{beamsplitter}
\end{equation}
The Kerr medium transformation in each arm is given by\cite{ref12a,ref12b}
\begin{equation}
        \hat{\cal S}_i (\chi; \tau) = \exp \left( -i \tau \hat{a}_i^{\dag} \hat{a}_i
        - i \chi_i \hat{a}_i^{\dag 2} \hat{a}_i^2 \right)
\label{kerrint}
\end{equation}
where $i=1,2$ and $\hat{a}_1 = \hat{a}$, $\hat{a}_2 = \hat{b}$
for transformation $\hat{\cal S}_i$, $i=1,2$.  The normally-ordered
interaction is employed rather than the symmetrically-ordered form also
found in the literature.  In eq.\ (\ref{kerrint}) the nonlinearity
coefficient $\chi_i$ is proportional to the nonlinear coefficient
$\chi^{(3)}$ of the medium and the interaction time within the medium.
The delay operator is
\begin{equation}
	\hat{\Delta}(\Delta) = \exp (i \Delta \hat{b}^{\dag} \hat{b})
\end{equation}
and introduces the linear phase shift which occurs between the arms of
the Mach-Zehnder interferometer.  For the Sagnac interferometer,
$\Delta=0$ and $\chi_1 = \chi_2$ is assumed.

The interferometer output state is $\hat{\cal I} |\alpha\rangle_a
|\beta\rangle_b$.  The beamsplitter transformation given in eq.\ 
(\ref{beamsplitter}) transforms the product coherent state
as follows
\begin{equation}
	\hat{\cal B}|\alpha\rangle_a |\beta\rangle_b =
	|2^{-1/2}(\alpha + i \beta) \rangle_1 
	|2^{-1/2}(\beta + i \alpha) \rangle_2
\label{beamsplitter2}
\end{equation}
for $1$ and $2$ the two beamsplitter output fields.  Thus the output
state is also a direct product of coherent states at its output given a
direct product at the input.  In fact this result can be generalised for
any semiclassical state.  A semiclassical state possesses a well-defined
positive-definite Glauber-Sudarshan P-representation\cite{ref9,ref13} which
behaves like a probability distribution on phase space.  A semiclassical
product state input $\hat{\rho}_a \otimes \hat{\rho}_b$, for
$\hat{\rho}_a$ the density matrix for state $a$ and similar for the
input state for $b$, can be expressed as
\begin{equation}
	\hat{\rho}_a \otimes \hat{\rho}_b =
	\left[ \int \frac{d^2 \alpha}{\pi} P_a (\alpha)|\alpha\rangle_a
	\langle\alpha| \right] \otimes \left[
	\int \frac{d^2 \beta}{\pi} P_b (\beta)
	|\beta\rangle_b \langle\beta| \right],
\end{equation}
where $P_a (\alpha)$ and $P_b (\beta)$ are the P-representations for the
states $\hat{\rho}_a$ and $\hat{\rho}_b$.  The output field from the
beamsplitter is given by
\begin{eqnarray}
	\hat{\cal B} \rho_a \otimes \rho_b \hat{\cal B}^{\dag} &=&
	\int \frac{d^2 \alpha}{\pi} P_a (\alpha) 
	\int \frac{d^2 \beta}{\pi} P_b (\beta) \nonumber \\ 
	& & | 2^{-1/2}( \alpha + i \beta ) \rangle_a
	\langle 2^{-1/2} (\alpha + i \beta ) |
	\otimes
	| 2^{-1/2} ( \beta + i \alpha ) \rangle_b
	\langle 2^{-1/2} ( \beta + i \alpha ) |
\label{beamout}
\end{eqnarray}
and a mixture of coherent states entering into a beamsplitter is 
transformed into
the obvious mixture of product coherent states at the output.  
For nonclassical fields this incoherent mixture of product states
does not hold as we shall see.

After the beam is split a path difference between the two arms can be
introduced, and this is represented mathematically 
by the delay operator $\hat{\Delta}$.
The delay operator acts on the product coherent state of eq.\ 
(\ref{beamsplitter2}) which leaves the beamsplitter to produce the state
\begin{equation}
	\hat{\Delta} \hat{\cal B} |\alpha\rangle_a |\beta\rangle_b =
	|2^{-1/2} (\alpha + i \beta)\rangle_1 
	|2^{-1/2} e^{i \Delta} (\beta + i \alpha)\rangle_2 .
\end{equation}
A phase shift of $\Delta$ has been effected in arm $2$ relative to arm $1$
of the interferometer.

The nonlinear Kerr transformation (\ref{kerrint}) transforms the coherent
state to\cite{ref12a,ref12b} 
\begin{equation}
	|\alpha \rangle^{\chi;\tau} \equiv \hat{\cal S}(\chi) |\alpha \rangle
	= \exp (-|\alpha |^2 /2) \sum^{\infty}_{n=0}
	\frac{(\alpha e^{i(\chi - \tau)})^n }{\sqrt{n!}}
	\exp (-i \chi n^2 )|n\rangle 
\label{kerrtransform}
\end{equation} 
which is henceforth referred to as a `sheared state,' a term which
describes the shearing of the Gaussian Q-function for the coherent
state over short times \cite{ref12a,ref12b}.  The rotating frame can be
chosen by setting $\tau =0$.  (Alternately the frame for which $\tau
= \chi$ is also used.)  The sheared state 
$|\alpha \rangle ^{\chi ; \tau =0}$ 
is a special case of the generalized coherent states of Titulaer
and Glauber \cite{ref14,ref15}, which can always be represented as a
continuous sum of coherent states \cite{ref15,stoler}.  Sheared 
states in particular have been discussed in this form
by Miranowicz {\it et al}.\ \cite{ref16} and by Gantsog and Tana\'{s}
\cite{ref17}, and can be expressed as the superposition
\begin{equation}
        |\alpha \rangle ^{\chi ;\tau =0} = \int_0^{2 \pi} 
        \frac{d \varphi}{2 \pi} f_{\chi }(\varphi)
        |\alpha e^{i (\chi - \varphi)} \rangle 
\label{alphainteg}
\end{equation} 
with
\begin{equation} 
        f_{\chi } (\varphi) = \sum_{n=0}^{\infty} \exp (in \varphi - i \chi n^2).
\end{equation} 
The phase function exhibits interesting properties and is discussed
further in Ref.\ \cite{ref18}.  For $\chi / \pi$ a rational number
the integral (\ref{alphainteg}) becomes a discrete sum over a finite
number of coherent states \cite{ref16}.

If $\chi / \pi $ is a rational number then there exists an integer
quantity $N$, such that
\begin{equation} 
	|\alpha \rangle ^{\chi; \tau =0} = 
	\sum_{n=1}^N c_n |\alpha e^{i \pi n/N} \rangle .
\label{alphasum}
\end{equation} 
This results because the factor
$\exp [i \chi n(n-1)]$ in eq.\ (\ref{kerrtransform}) is 
periodic when $\chi / \pi $ is rational.
For $r,s$ integers which are relatively prime and $\chi = 2(r/s)
\pi$, we observe that 
\begin{equation} 
        \frac{r}{s} n^2 = \frac{r}{s} (n+N)^2 \mbox{ mod } 1
\label{pqrelprime}
\end{equation} 
for $N=s$.  If $s$ is not prime, then $N \leq s$ is possible; for
example eq.\ (\ref{pqrelprime}) is satisfied by $N=s/2$ for $s$ a
multiple of $4$ \cite{ref18}.  For the special case that 
$\chi = \pi /2$ we have $r=1$, $s=4$, and $N=2$ and we find that
\begin{equation} 
	|\alpha \rangle ^{\chi = \pi /2; \tau =0} = 2^{-1/2}
	\left( e^{- i \pi /4} |i \alpha \rangle + e^{i \pi /4}|-i \alpha
	\rangle \right) .
\label{schrodcat1}
\end{equation} 
This superposition state\cite{ref12a,ref20} has been discussed in the
context of optical analogs to Schr\"{o}dinger's cat state
\cite{ecs0,ref21,ref22a,ref22b,ref22c}.  Similar analyses can 
yield a superposition of phase
states \cite{phase}.
More generally the coefficients of the state
(\ref{alphasum}) are determined by solving the $N$ simultaneous
equations \cite{ref15}
\begin{equation} 
	\sum_{n=1}^N c_n e^{2i \pi kn/N} = e^{i \chi k(k-1)}
\end{equation} 
for $k=0,1,...,N-1$.  Using the method of Gantsog and Tana\'{s}
\cite{ref17}, this can be solved to determine that
\begin{equation} 
	c_n = \frac{1}{N} \sum_{k=0}^{N-1} 
	\exp [-i 2 \pi kn/N - i \chi k(k-1)],
\label{cncalc}
\end{equation} 
where $n = 1,2,...,N$.

The output field of the interferometer is given by
\begin{equation} 
	\hat{\cal I} |\alpha \rangle_a |\beta \rangle_b =
	\hat{\cal B}
	|2^{-1/2}(\alpha +i \beta ) \rangle_1^{\chi_1}
	|2^{-1/2} e^{i \Delta}(\beta  +i \alpha  ) \rangle_2^{\chi_2} .
\end{equation} 
If the states $|2^{-1/2}(\alpha +i \beta ) \rangle_1^{\chi_1}$ and
$|2^{-1/2} e^{i \Delta}(\beta  +i \alpha  ) \rangle_2^{\chi_2}$ are
semiclassical then the output could be be written as a product
coherent state or a mixture of product coherent states in the way
that eq.\ (\ref{beamout}) is written.  However the sheared states,
despite being generalised coherent states, are not semiclassical
states.  The nature of the interferometer output states are
considered in the next section.

\section{Output states}

In order to analyze the output states of the interferometer, the
coherent field with amplitude $\beta $ is now restricted to the
vacuum state by setting $\beta =0$.  Thus the output state that we
wish to consider is given by the formal expression
\begin{equation} 
	\hat{\cal I} |\alpha \rangle_a |0 \rangle_b = \hat{\cal B} |\alpha / \sqrt{2}
	\rangle_1^{\chi_1} |i \alpha /\sqrt{2} \rangle _2^{\chi _2}.
\label{precalcgeneraloutput}
\end{equation} 
The sheared state can be expressed as a superposition of coherent
states according to expression (\ref{kerrtransform}).  By
substituting this result into eq.\ (\ref{precalcgeneraloutput}), we
obtain the formal result 
\begin{eqnarray} 
	\hat{\cal I}|\alpha \rangle_a |0 \rangle_b &=& 
        \int_0^{2 \pi} \frac{d \varphi_1}{2 \pi } f_{\chi_1} (\varphi_1)
	\int_0^{2 \pi} \frac{d \varphi_2}{2 \pi } f_{\chi_2} (\varphi_2)
	\nonumber \\
	&& 
	|\frac{1}{2} \alpha (e^{i(\chi_1 - \varphi_1)}
	- e^{i(\Delta + \chi_2 - \varphi_2)}) \rangle_{a'}
	|\frac{1}{2}i \alpha 
	(e^{i(\Delta + \chi_2 - \varphi_2)}
	+ e^{i(\chi_1 - \varphi_1)} \rangle_{b'} .
\label{twomodeproduct}
\end{eqnarray}
The output state is a superposition of two-mode product coherent states.

The simplest case arises for the linear interferometer for which
$\chi_1 =0 = \chi_2$.  In this case we can show that
\begin{equation} 
	\hat{\cal I}|\alpha \rangle_a |0 \rangle_b =
	|\alpha (1-e^{i \Delta})/2 \rangle_{a'}
	|i \alpha (1+e^{i \Delta})/2 \rangle_{b'}
\end{equation} 
as expected.  The output state is unchanged for $\chi_1 = \pi =
\chi_2$ and both $\chi _1=0= \chi _2- \pi $ and 
$\chi _1 - \pi =0= \chi _2$.  A periodic behaviour is evident in
$\chi_1 - \chi_2$ parameter space.

The case for which $\chi_1 = \pi/2$ and $\chi_2 =0$ is interesting as
well.  In this case we find that
\begin{eqnarray} 
	\hat{\cal I} |\alpha \rangle_a |0 \rangle_b
        &=& 2^{-1/2} e^{-i \pi/4} [|(\alpha (i-e^{i \Delta})/2 \rangle_{a'}
	|i \alpha (i+e^{i \Delta})/2 \rangle_{b'} \nonumber \\
	&& + i |- \alpha (i+e^{i \Delta})/2 \rangle_{a'}
	|-i \alpha (i-e^{i \Delta})/2 \rangle_{b'}].
\label{generaloutputch1}
\end{eqnarray} 
For $\Delta = \pi /2$, the entangled coherent state 
\cite{ecs0,ecs1,ecs2,ecs3,ecs4,ecs5,ecs6,ecs7,ecs8}
\begin{equation} 
	\hat{\cal I} |\alpha \rangle_a |0 \rangle_b =
	2^{-1/2} e^{-i \pi /4}
	[|0 \rangle_{a'} |\alpha \rangle_{b'}
	+i |-i \alpha \rangle_{a'} |0 \rangle_{b'} ]
\end{equation} 
is obtained.  However, $\Delta \neq 0$ and therefore this state is not
obtained by a Sagnac interferometer in contrast to the entangled
state of Ref.\ \cite{ecs1}.  
The reason for this difference is the
normal ordering of the nonlinear interaction here as opposed to the
symmetric ordering used in 
Ref.\ \cite{ecs1}.  Physically alternate
orderings introduce different linear phase shifts.

In fact the more general state (\ref{generaloutputch1}) can be
regarded as entangled as well.  A superposition of two mode product
coherent states
\begin{equation} 
	| \alpha_1 \rangle_a |\beta_1 \rangle_b
	+ e^{i \varphi}|\alpha_2 \rangle_a |\beta_2 \rangle_b
\label{ecs1}
\end{equation} 
is entangled provided that the inner products 
$|_a \langle \alpha_1|\alpha_2 \rangle_a |$ and
$|_b \langle \beta_1|\beta_2 \rangle_b |$ are sufficiently small.  As the 
overlap functions for the $a$ and $b$ states of expression
(\ref{generaloutputch1}) are given by
\begin{eqnarray}  
	_{a'} \langle \alpha (i-e^{i \Delta })/2|
	- \alpha (i +e^{i \Delta })/2 \rangle _{a'}
	&=& \exp (-| \alpha |^2 [1-i \cos \Delta ]/2 ) \nonumber \\
	&=& _{b'} \langle i \alpha (i +e^{i \Delta })/2
	|-i \alpha (i-e^{i \Delta })/2 \rangle _{b'},
\end{eqnarray} 
the inner products quickly become small as $|\alpha |^2 \rightarrow
\infty$.  Consequently the state (\ref{generaloutputch1}) satisfies
the criteria for being an entangled coherent state for all $\Delta $.
Thus, although the output state for $\Delta =0$ differs from that of
Ref.\ \cite{ecs1}, the output is 
nevertheless an entangled coherent state.  

Another special case for the interferometer arises for $\chi \equiv 
\chi_1 = \chi_2$ and $\Delta =0$.  This restriction corresponds to
the case generally used in squeezed light experiments
\cite{ref4}.  The output state is given by
\begin{eqnarray} 
	\hat{\cal I} |\alpha \rangle_a |0 \rangle_b &=&
	\int_0^{2 \pi } \frac{d \varphi }{2 \pi } f_{\chi }( \varphi )
	\int_0^{2 \pi } \frac{d \varphi' }{2 \pi } f_{\chi' }( \varphi )
	\nonumber \\
	&& 
	|2^{-1} \alpha e^{i \chi }
	(e^{- i \varphi } - e^{-i \varphi '} \rangle_{a'}
	|2^{-1} i 
	(e^{- i \varphi' } + e^{-i \varphi } \rangle_{b'}.
\end{eqnarray} 
For the case that $\chi = \pi /2$, we have
\begin{equation} 
	\hat{\cal I} |\alpha \rangle _a |0 \rangle _b
	= \frac{1}{2} 
	(|\alpha \rangle _{a'} + |- \alpha \rangle _{a'})
	|0 \rangle _{b'}
	+\frac{1}{2} i |0 \rangle_{a'}
	(|i \alpha \rangle _{b'} + |-i \alpha \rangle _{b'}).
\label{schrodcat2}
\end{equation} 
This state corresponds to an entanglement of a Schr\"{o}dinger cat
state from port $a'$ and a vacuum at port $b'$ with a vacuum state
from port $a'$ and a Schr\"{o}dinger cat state from port $b'$.
However the Schr\"{o}dinger cat state in expression (\ref{schrodcat2})
is very different from the Schr\"{o}dinger cat state in eq.\
(\ref{schrodcat1}).  This difference is most evident in the photon
number distribution.  The photon number distribution of eq.\
(\ref{schrodcat1}) is identical to the distribution of the coherent
state $| \alpha \rangle $, but the photon number distribution of the state
\begin{equation} 
	[2(1+ e^{ -2 | \alpha |^2})]^{-1/2}
	(|\alpha \rangle + |- \alpha \rangle )
	= \cosh (|\alpha |^2)
	\sum_{n=0}^{\infty} \frac{\alpha ^{2n}}{\sqrt{(2n)!}}
	|2n \rangle 
\end{equation} 
is quite different and is a superposition of even photon number
states only: hence the nomenclature `even coherent states'
\cite{ecs0,ref22a,ref22b,ref22c}.

Other interesting features arise for various values of $\chi _1, \chi
_2$, and $\Delta$, but the interesting states are special cases of
eq.\ (\ref{twomodeproduct}).  One of these special cases 
arises for $\chi_1 / \pi $  and $\chi_2 / \pi$ rational 
numbers $2p_1/q_1$  and $2p_2/q_2$, for each pair $p_1$, $q_1$ 
and $p_2$, $q_2$ relatively prime integers
which are very small.  Under this condition the sheared state is a
superposition of very few distinguishable coherent states according
to the sum (\ref{alphasum}).  That is, for the nonlinearities
$\chi_1$ and $\chi_2$, there exist integers $M$ and $N$ such that
\begin{eqnarray}
	|\alpha \rangle ^{\chi_1,\tau=0}
	= \sum_{m=1}^M c_m |\alpha e^{i \pi m/M} \rangle 
	\label{alphasum1} ,
	\\
	|\alpha \rangle ^{\chi_2,\tau=0}
	= \sum_{n=1}^N c_n |\alpha e^{i \pi n/N} \rangle 
	\label{alphasum2} .
\end{eqnarray}
By substituting (\ref{alphasum1}) and (\ref{alphasum2})
into the interferometer equation in (\ref{precalcgeneraloutput}),
the interferometer output state is found to be
\begin{equation}
	\hat{\cal I} |\alpha \rangle _a |0 \rangle _b
	= \sum_{m=1}^M \sum_{n=1}^N c_m c_n
	|\alpha (e^{i \pi m/M} - e^{i \pi n/N})/2 \rangle _{a'}
	|i \alpha (e^{i \pi m/M} + e^{i \pi n/N})/2 \rangle _{b'} ,
\end{equation}
where the coefficients $c_m$ and $c_n$ can be calculated from 
the application of (\ref{cncalc}).
This shows that the general output state for $\chi_1$ and $\chi_2$
rational is an entangled coherent state with a finite cardinality
for the basis set.

The other interesting parameter regime
for $\chi $ corresponds to $\chi / \pi $ a small quantity.  This case
is important in the squeezed light experiments and is the subject of
the next section.

\section{Weak nonlinearities and squeezing}

The weakly nonlinear interferometer is used for squeezed light
\cite{ref24} experiments 
and corresponds to small to moderate lengths of nonlinear
material in each arm of the interferometer.  The
quantity $\chi $ can be set to a very small number.  Here we wish to
see how the formal results established in the previous sections can
be used to understand the weakly nonlinear interferometer and the
phenomenon of squeezing.

In order to understand the weakly nonlinear Mach-Zehnder
interferometer, we must understand the sheared state $|\alpha \rangle
^{\chi ; \tau }$ for which $\chi $ is small.  The sheared state can
be expressed as
\begin{equation} 
	|\alpha \rangle ^{\chi ; 0 } \equiv
	\hat{\cal S}(\chi; 0 ) \hat{D}(\alpha )|0 \rangle 
\label{shearstate}
\end{equation} 
for $\hat{\cal S}(\chi ;0 )$ the shear operator (\ref{kerrint}) and 
$\hat{D}(\alpha )$ the displacement operator (\ref{displacement}).
The unitary operators can be rewritten as
\begin{equation} 
	\hat{\cal S}(\chi ;0)\hat{D}(\alpha ) = 
	\hat{D}(\alpha ) \exp [-i \chi (\hat{a}^{\dagger }
	+ \alpha^{*})^2 (\hat{a}+\alpha )^2].
\label{kerrdisplace}
\end{equation} 
Suppose that the photon number $|\alpha |^2 \rightarrow \infty$ and the
nonlinear parameter $\chi \rightarrow 0$ such that
$\eta = \chi \alpha ^2$.
Consequently terms in 
eq.\ (\ref{kerrdisplace}) with
coefficients of order $\eta / |\alpha |$ and smaller
are negligible.  Relegating details of the calculation to
Appendix A, the state (\ref{shearstate}) can be approximated by
\begin{equation}
	|\alpha \rangle ^{\chi ; 0}
	\approx \exp(-i \Lambda) \hat{D}(\alpha + \delta)
	  \hat{S}(\varepsilon ) \hat{S}(-e^{2i \sigma} \varepsilon )|0 \rangle,
\label{shearedstateapprox2}
\end{equation}
where $\hat{S} (\varepsilon )$ is the squeeze operator 
\begin{equation}
	\hat{S}(\varepsilon ) = \exp [(\varepsilon ^* \hat{a}^2 
	- \varepsilon  \hat{a}^{\dagger 2})/2] ,
\label{squeezeoperator}
\end{equation}
$\hat{D}(\rho)$ is the displacement operator (\ref{displacement}), 
and $\delta$ and $\Lambda$ are complex functions of $\alpha$
and $\chi$ given by eqns.\ (\ref{delta}) and (\ref{lambda}) respectively.

It is evident that the state (\ref{shearedstateapprox2})
is a vacuum state which has been squeezed along two different
axes, then displaced.

The output state for the weakly nonlinear Mach-Zehnder interferometer
is given by (\ref{precalcgeneraloutput}), which, by
using eq.\ (\ref{shearedstateapprox2}), can be approximated by
\begin{eqnarray} 
	\hat{\cal I}|\alpha \rangle _a |0 \rangle _b  
	& \approx & e^{-i (\Lambda_1 + \Lambda_2)} \hat{\cal B} 
	\hat{D}_1 (\omega_1 + \delta_1)
	\hat{D}_2 (\omega_2 + \delta_2 ) \nonumber \\
	&& \hat{S}_1 (\varepsilon _1) 
	\hat{S}_1 ( -e^{2i \sigma_1} \varepsilon _1)
	\hat{S}_2 (\varepsilon _2)
	\hat{S}_2 (-e^{2i \sigma_2} \varepsilon _2) 
	|0 \rangle _1 |0 \rangle _2 .
\label{squeezeintout}
\end{eqnarray} 
where $\omega_1 = \alpha / \sqrt{2}$ and 
$\omega_2 = i \alpha / \sqrt{2}$, and
$\Lambda_i$, $\delta_i$, $\varepsilon _i$, and $\sigma_i$
(for $i = 1,2$) are complex functions of 
$\alpha $, $\chi_1$, and $\chi_2$ which are given by eqs.\ 
(\ref{lambdai}), (\ref{deltai}), (\ref{varepsiloni}), and
(\ref{sigmai}) respectively in Appendix B.

If $\chi \equiv \chi_1 = \chi_2$, then $\varepsilon _2 = - \varepsilon _1$ and
$\sigma _2 = \sigma _1$.  In this case, eq.\ (\ref{squeezeintout})
can be calculated to find that
\begin{eqnarray}
	\hat{\cal I} |\alpha \rangle _a |0 \rangle _b & \approx &
	e^{-i(\Lambda_1 + \Lambda_2)}
	\hat{D}_{a'} (\gamma_1) \hat{D}_{b'} (\gamma_2) \nonumber \\
	&& \hat{S}_{a'}(\varepsilon _1) 
	\hat{S}_{a'} (-e^{2i \sigma_1} \varepsilon _1)
	\hat{S}_{b'} (- \varepsilon _1) 
	\hat{S}_{b'} (e^{2i \sigma_1} \varepsilon _1)
	|0 \rangle_{a'} |0 \rangle_{b'} 
	\label{squeezeintout2}
\end{eqnarray}
where $\gamma_i$ 
is a complex function of $\alpha $ and $\chi$ and
is given by eq.\ (\ref{gammai}) (for $i = 1,2$).

Thus, when $\chi_1 = \chi_2$, the output state is a product state of the
squeezed coherent state at port $a'$ and an orthogonally squeezed coherent
state at port $b'$.  If we adopt assumptions about strong coherent
fields and weak nonlinearities then the treatment of squeezed
coherent states from each port is valid.

It is interesting to note that 
for $\chi_1 = \chi_2$ the two coherent states enter the two
input ports of the interferometer and exit again as two squeezed
states.  The output state is a product state as well.

\section{Two-field interaction and Entangled Coherent States}

In Section III, it was shown that the nonlinear Mach-Zehnder
interferometer with coherent state
inputs in general results in an output of entangled coherent states.
Here, it is shown that entangled coherent states can also be created
using only an ideal Kerr nonlinearity with two coherent state inputs, 
without the need for an interferometer. 

The Kerr transformation for a single field input was given by eq.\
(\ref{kerrint}).  When two input fields, $1$ and $2$, simultaneously 
enter into the Kerr cell, the Kerr transformation is given by \cite{qndm,qndm2}
\begin{equation} 
	\hat{\cal S}_{12}( \chi , \tau ) = \exp [-i \tau 
	(a_1^{\dagger }a_1 + a_2^{\dagger }a_2) - i \chi
	(a_1^{\dagger 2}a_1^2 + a_2^{\dagger 2}a_2^2
	+ 4a_1^{\dagger }a_1 a_2^{\dagger }a_2 )] .
\label{kerrop2}
\end{equation} 
The term $a_1^{\dagger}a_1 a_2^{\dagger}a_2$ in the exponential represents
the nonlinear two-field interaction which occurs 
where the two input fields superpose in the nonlinear cell.  The effect of 
this is a phase-shift dependent on the photon numbers of both fields,
which leaves the photon number of each field unchanged.  This
property enables the two-field nonlinear interaction to be used as a 
quantum non-demolition measurement of photon number 
\cite{qndm,qndm2,qndmcomp}, 
where one field provides the signal and the other field is used for
the measurement.

A simple form of the entangled coherent state can be obtained by
using three nonlinear cells, one for the interaction, preceded by two to cancel
the shearing effect on the state in phase space without cancelling the
interaction term, as shown in Fig.\ 2.  
(The more complicated result when only one nonlinear cell is used 
is calculated in Appendix C.)  The two input coherent states, injected
into the pair of nonlinear cells, could in fact
be created by directing a single coherent beam into a beamsplitter with a 
product coherent state output.

In this type of arrangement, the total operator 
of eq.\ (\ref{kerrop2}) is then reduced to the form 
\begin{equation} 
	\hat{\cal S}'_{12} = \exp (-4 i \chi \hat{a}_1^{\dag} \hat{a}_1
        \hat{a}_2^{\dag} \hat{a}_2 ).
\end{equation} 
For two coherent state inputs, the output in the Fock state basis
is
\begin{eqnarray} 
	|\alpha , \beta \rangle_{12}^{\chi , 0}
	& \equiv & \hat{\cal S}' (\chi , 0)
	\hat{D}_1 (\alpha ) \hat{D}_2 (\beta )
	|0 \rangle _1 |0 \rangle _2  \nonumber \\
        &=& e^{-(|\alpha |^2 + |\beta |^2)/2}
        \sum_{m=0}^{\infty} \sum_{n=0}^{\infty}
	\frac{\alpha ^m \beta ^n}{\sqrt{m! n!}}
	e^{-4i \chi mn} |m \rangle _1 |n \rangle _2 
\label{redkerrout}
\end{eqnarray} 
which is a generalization of expression (\ref{shearstate}).  This
can also be expressed in the coherent state basis as
\begin{equation}  
	|\alpha , \beta \rangle_{12}^{\chi ; \tau =0}
	= \int_0^{2 \pi } \frac{d \theta }{2 \pi }
	\int_0^{2 \pi } \frac{d \varphi }{2 \pi }
	g'_{\chi } (\theta , \varphi )
	|e^{-i \theta } \alpha \rangle _1
	|e^{-i \varphi } \beta \rangle _2 ,
\label{gentwofieldkerrout}
\end{equation} 
with
\begin{equation} 
	g'_{\chi } (\theta , \varphi )
	= \sum_{p,q=0}^{\infty} 
	\exp (i \theta p + i \varphi q - 4 i \chi pq) .
\end{equation} 
The output state (\ref{gentwofieldkerrout}) 
is an entangled coherent state.

If $\chi / \pi $ is a rational number $2r/s$, a finite
entangled sum of coherent states results:
\begin{equation} 
	|\alpha , \beta \rangle ^{\chi ; 0}
	= \sum_{m=1}^N \sum_{n=1}^N c_{mn} 
	|e^{2 \pi i m/N} \alpha \rangle _1 
	|e^{2 \pi i n/N} \beta \rangle _2 ,
\label{entangledoutsimple}
\end{equation} 
where $N=s$ if $r$ and $s$ are relatively prime, and $N<s$ is
possible otherwise.
The coefficients $c_{mn}$ are found by solving the $N^2$ simultaneous
equations
\begin{equation} 
	\sum_{m=1}^N \sum_{n=1}^N c_{mn} e^{2 \pi i km/N}
        e^{2 \pi i ln/N} = e^{-4i \chi mn} .
\end{equation} 
Solving these equations using an extension of the method of Gantsog 
and Tana\'{s} \cite{ref17} gives the result
\begin{equation} 
	c_{mn} = \frac{1}{N^2} \sum_{k=0}^{N-1} \sum_{l=0}^{N-1}
	\exp [- 2 i (\pi km/N + \pi ln/N + 2 \chi kl )] .
\end{equation} 
The factor $\exp(-4 i \chi kl)$ in the above expression 
is the nonlinear interaction term.  The presence of this factor 
means that the general output is an entangled sum of coherent states, 
unless  $2 \chi / \pi $ is an integer.  In the latter case, 
$\exp(-4 i \chi kl) =1$ and the output will be a product state.

For $\chi = \pi /4$, the resulting output state is 
\begin{eqnarray}  
	|\alpha , \beta \rangle ^{\chi ; 0}
	&=& \frac{1}{2} [ |\alpha \rangle _1
	(|\beta \rangle _2 + |- \beta \rangle _2 )
	+ |- \alpha \rangle _1 
	(|\beta \rangle _2 - |- \beta \rangle _2 )]
        \nonumber \\
	&=& \frac{1}{2} [(|\alpha \rangle _1 + |-\alpha \rangle _1)
	|\beta \rangle _2
	+ (|\alpha \rangle _1 - |- \alpha \rangle _1)
	|-\beta \rangle _2 ].
\label{kerronlyecs}
\end{eqnarray} 
This is an entangled state, comparable to those in eqs.\ (\ref{ecs1}) 
and (\ref{schrodcat2}).

A difference between the entangled coherent state in
eq.\ (\ref{kerronlyecs}) and the entangled
coherent states (\ref{ecs1}) from the nonlinear interferometer
can be seen if one of the input states in eq.\ (\ref{kerronlyecs})
is in the vacuum state.  If we set $\beta =0$, then the output
for the Kerr cell becomes
\begin{equation} 
	|\alpha , 0 \rangle_{12}^{\chi ; 0} 
	= |\alpha \rangle _1 |0 \rangle _2
\end{equation} 
Unlike the entangled coherent states produced in the nonlinear
interferometer in eqs.\ (\ref{ecs1}) and
(\ref{schrodcat2}), for a single Kerr cell an entangled coherent
state only results when both inputs are not in the vacuum state.

There are a number of advantanges to this new alternative approach to
creating entangled coherent states.  An interferometer uses nonlinear
cells, mirrors, and beam splitters.  The approach here, using two
coherent inputs into a nonlinear cell, produces entangled coherent
states without the need for mirrors or beamsplitters.  Thus, many
technical difficulties of interferometry are eliminated.

\section{Conclusion}

The formalism which has been presented here has clarified 
that for coherent state inputs, the general output of the
nonlinear Mach-Zehnder interferometer consists of 
entangled coherent states.  For weak nonlinear evolution, a
squeezed state output results.  At the other extreme of high
values of the nonlinear Kerr coefficient,
$\chi = \pi /2$, the entangled coherent state 
$2^{-1/2} (|0 \rangle _a |\alpha \rangle _b 
+ i |-i \alpha \rangle _a |0 \rangle _b )$ results for a single
coherent state input into one port of the interferometer, and
a vacuum state entering the other port.  For states
in between these two extremes, in general a type of entangled
coherent state will be produced.  

It has also been demonstrated that entangled coherent states can
also be produced using only an ideal Kerr nonlinearity without the
need for an interferometer.  For two coherent input states
$|\alpha \rangle $ and $|\beta \rangle $ into an ideal Kerr
nonlinearity, the interaction between the two states produces
the entangled coherent state output 
$2^{-1} [|\alpha \rangle _a (|\beta \rangle _b + |- \beta \rangle _b)
+ |- \alpha \rangle _a (|\beta \rangle _b - |- \beta \rangle _b)]$.
While still an entangled state, this entangled state produced by
a nonlinear Kerr cell differs from that produced by a nonlinear
Mach-Zehnder interferometer since in the former case, both inputs
must not be in the vacuum state.  If one of the input states is the
vacuum state, the other coherent state input passes through unchanged.

\appendix

\section{Approximating the Displacement-Shear Unitary Operator}

Recall from eq.\ (\ref{kerrdisplace}) that
\begin{equation} 
	\hat{\cal S}(\chi ;0)\hat{D}(\alpha ) = 
	\hat{D}(\alpha ) \exp [-i \chi (\hat{a}^{\dagger }
	+ \alpha^{*})^2 (\hat{a}+\alpha )^2],
\label{kerrdisplace2}
\end{equation} 
and that we have introduced the quantity
\begin{equation}
	\eta = \chi \alpha^2 ,
\end{equation}
and we allow $\chi \rightarrow 0$ and $|\alpha|^2 \rightarrow \infty$
such that $\eta$ remains constant.

Expanding the exponential in (\ref{kerrdisplace2}) and keeping only
those $\alpha$ terms of order 2 or greater produces the result
\begin{eqnarray}
	\hat{\cal S}(\chi ;0)\hat{D}(\alpha ) \approx 
	\hat{D}(\alpha ) 
	\exp (&&  -4 i |\eta|^2 \hat{a}^{\dagger} \hat{a}
	- 2 i \alpha ^* \eta \hat{a}^{\dagger}
	- 2 i \alpha \eta^* \hat{a} \nonumber \\
	&& 
	- i \eta \hat{a}^{\dagger 2}
	- i \eta ^* \hat{a}^2
	- i \chi |\alpha |^4 ).
\label{appa1}
\end{eqnarray}

On the other hand, the term 
$\hat{S}(\varepsilon )
\hat{D}(\alpha ) \hat{R}(\rho) \hat{D}^{\dagger}(\alpha )
\hat{S}^{\dagger}(\varepsilon )$ 
can be expanded so that
\begin{eqnarray}
	&& \hat{S}(\varepsilon )
	\hat{D}(\alpha ) \hat{R}(\rho) \hat{D}^{\dagger}(\alpha )
	\hat{S}^{\dagger}(\varepsilon ) \nonumber \\
	&& = \exp [i \sigma (\hat{a}^{\dagger} \cosh |\varepsilon |
	+ \hat{a} \frac{\varepsilon }{|\varepsilon |}
	\sinh |\varepsilon | - \alpha ^* )
	(\hat{a} \cosh |\varepsilon | + \hat{a}^{\dagger}
	\frac{\varepsilon ^*}{|\varepsilon |} \sinh |\varepsilon |
	- \alpha )] .
\label{appa2}
\end{eqnarray}
Therefore, (\ref{appa1}) can be re-expressed as 
\begin{equation}  
	|\alpha \rangle ^{\chi  ; 0 }  \approx  
	\exp [i \chi |\alpha |^4 - i \sigma (\sinh ^2 |\varepsilon | 
	+ |\alpha |^2 ) ]
	\hat{D} (\alpha ) \hat{S}(\varepsilon ) \hat{D} (\rho )
	\hat{R} (\sigma ) \hat{D}^{\dagger} (\rho ) 
	\hat{S}^{\dagger} (\varepsilon ) |0 \rangle ,
\label{shearedstateapprox}
\end{equation} 
where 
$\hat{D}(\rho)$ is the displacement operator (\ref{displacement}), 
$\hat{S} (\varepsilon )$ is the squeeze operator
(\ref{squeezeoperator}), 
and $\hat{R}(\sigma)$ is the rotation operator
$\hat{R}(\sigma) = \exp (i \sigma \hat{a}^{\dagger} \hat{a})$,
as long as the following simultaneous equations hold:
\begin{equation}
	\sigma \cosh 2 |\varepsilon | = -4 |\eta|^2 
\end{equation}
\begin{equation}
	\sigma \frac{\varepsilon }{|\varepsilon |} 
	\sinh 2|\varepsilon | = - 2 \eta^* 
\end{equation}
and
\begin{equation}
	\sigma (- \rho ^* \cosh |\varepsilon |
	- \rho \frac{\varepsilon }{|\varepsilon |}
	\sinh |\varepsilon | ) = - 2 \alpha \eta^* .
\end{equation}
Solving these equations gives the resultant expressions 
\begin{equation}
	\rho = 
		\frac{2 \alpha ^* \eta }{\sigma} \cosh |\varepsilon |
	       - \frac{2 \alpha \eta^*}{\sigma} 
		\frac{\varepsilon ^*}{|\varepsilon|}
		\sinh |\varepsilon | ;
	\label{rho}
\end{equation}
\begin{equation}
	\varepsilon = \frac{1}{2} \frac{\eta ^*}{|\eta|} \tanh^{-1} 
	\left( \frac{1}{2|\eta|} \right); \label{varepsilon} 
\end{equation}
and
\begin{equation}
	\sigma = -4 |\eta|^2 \sqrt{1 - \frac{1}{4 |\eta|^2}} .
	\label{sigma} 
\end{equation}

Eq.\ (\ref{shearedstateapprox}) can be 
further simplified to obtain the result
\begin{equation}
	|\alpha \rangle ^{\chi ; 0}
	\approx \exp(-i \Lambda) \hat{D}(\alpha + \delta)
	  \hat{S}(\varepsilon ) \hat{S}(-e^{2i \sigma} \varepsilon )|0 \rangle,
\label{shearedstateapprox3}
\end{equation}
where $\delta$ and $\Lambda$ are given
by 
\begin{equation}
	\delta = \cosh |\varepsilon | \rho (1 - e^{i \sigma})
	- (\varepsilon / |\varepsilon |) \sinh |\varepsilon | \rho^{*}
	(1 - e^{- i \sigma } );
	\label{delta}
\end{equation}
and 
\begin{equation}
	\Lambda	= \chi |\alpha |^4 + \sigma (\sinh ^2 |\varepsilon |
	+ |\alpha |^2) + |\rho|^2 \sin \sigma - \mbox{Im} 
	\{ \alpha \delta^{*} \}.
	\label{lambda}
\end{equation}
The expression (\ref{shearedstateapprox3}) is obtained
by using the relation
\begin{equation}
	\hat{R} (\sigma ) \hat{D}^{\dagger} (\rho ) 
	\hat{S}^{\dagger} (\varepsilon )
	=  \hat{D}^{\dagger} (e^{i \sigma} \rho ) 
	\hat{S}^{\dagger} (e^{2i \sigma} \varepsilon ) 
	\hat{R} (\sigma )
\end{equation}
as well as the property for the displacement operator \cite{wallsmilburn}
\begin{equation}
	\hat{D}(\alpha) \hat{D}(\beta)
	= \hat{D} (\alpha + \beta )
	\exp (i \mbox{Im} \{ \alpha \beta ^* \} )
\end{equation}
and the commutation relation for $\hat{D}$ and $\hat{S}$
\cite{wallsmilburn} .

\section{The Output of the Nonlinear Interferometer with a
Weak Nonlinearity}

The output state for the weakly nonlinear Mach-Zehnder interferometer
was given in eq.\ (\ref{squeezeintout}), which was
\begin{eqnarray} 
	\hat{\cal I}|\alpha \rangle _a |0 \rangle _b  
	& \approx & e^{-i (\Lambda_1 + \Lambda_2)} \hat{\cal B} 
	\hat{D}_1 (\omega_1 + \delta_1)
	\hat{D}_2 (\omega_2 + \delta_2 ) \nonumber \\
	&& \hat{S}_1 (\varepsilon _1) 
	\hat{S}_1 ( -e^{i \sigma_1} \varepsilon _1)
	\hat{S}_2 (\varepsilon _2)
	\hat{S}_2 (-e^{2i \sigma_2} \varepsilon _2) 
	|0 \rangle _1 |0 \rangle _2 .
\label{squeezeintout3}
\end{eqnarray} 
In this equation, $\omega_1$ and $\omega_2$ are given by
\begin{eqnarray}
	\omega_1 = \alpha / \sqrt{2} , \label{omega1} \\
	\omega_2 = i \alpha / \sqrt{2} , \label{omega2}
\end{eqnarray}
and $\Lambda_i$ and $\delta_i$ are given by
\begin{equation}
	\Lambda_i = \chi |\omega_i|^4 + \sigma _i
	(\sinh^2 |\varepsilon _i| + |\omega_i|^2 )
	+ |\rho_i|^2 \sin \sigma _i 
	- \mbox{Im} \{ \omega_i \delta_i^* \}, \label{lambdai}  
\end{equation}
\begin{equation}
	\delta_i = \cosh|\varepsilon _i| \rho_i (1-e^{i \sigma _i})
	- (\varepsilon _i / |\varepsilon _i |)
	\sinh |\varepsilon _i| \rho^*_i (1-e^{-i \sigma _i}) ,
	\label{deltai} 
\end{equation}
with $\rho_i$ given by
\begin{equation}
	\rho_i = 
		\frac{2 \omega_i ^* \eta_i }{\sigma_i} \cosh |\varepsilon_i |
	       - \frac{2 \omega_i \eta^*_i}{\sigma_i} 
		\frac{\varepsilon ^*_i}{|\varepsilon_i|}
		\sinh |\varepsilon_i | ,
	\label{rhoi} 
\end{equation}
and $\varepsilon _i$ and $\sigma_i$ are given by
\begin{equation}
	\varepsilon_i = \frac{1}{2} \frac{\eta ^*_i}{|\eta_i|} \tanh^{-1} 
	(\frac{1}{2|\eta_i|}) , \label{varepsiloni} 
\end{equation}
\begin{equation}
	\sigma_i = -4 |\eta_i|^2 \sqrt{1 - \frac{1}{4 |\eta_i|^2}} ,
	\label{sigmai} 
\end{equation}
with $\eta_i$ given by
\begin{equation}
	\eta_i = \chi_i \omega_i^2 
	\label{etai}
\end{equation}
for $i=1,2$.

If $\chi \equiv \chi_1 = \chi_2$, then 
eq.\ (\ref{squeezeintout3}) can be calculated to obtain
the result given in eq.\ (\ref{squeezeintout2}), which was
\begin{eqnarray}
	\hat{\cal I} |\alpha \rangle _a |0 \rangle _b & \approx &
	e^{-i(\Lambda_1 + \Lambda_2)}
	\hat{D}_{a'} (\gamma_1) \hat{D}_{b'} (\gamma_2) \nonumber \\
	&& \hat{S}_{a'}(\varepsilon _1) 
	\hat{S}_{a'} (-e^{i \sigma_1} \varepsilon _1)
	\hat{S}_{b'} (- \varepsilon _1) 
	\hat{S}_{b'} (e^{2i \sigma_1} \varepsilon _1)
	|0 \rangle_{a'} |0 \rangle_{b'} .
	\label{squeezeintout4}
\end{eqnarray}
In eq.\ (\ref{squeezeintout4}), $\gamma_i$ 
is a complex function of $\alpha $ and $\chi$ which
is given by 
\begin{equation}
	\gamma_i = \frac{1}{\sqrt{2}} 
	\left[ \cosh |\varepsilon _1| \Gamma_i
	- \frac{\varepsilon _1}{| \varepsilon _1|} \Gamma_i^* \right] ,
	\label{gammai}
\end{equation}
for $i=1,2$, with 
\begin{eqnarray}
	\Gamma_1 &=& [ \cosh |\varepsilon _1| (\omega_1 + \delta_1)
	+ (\varepsilon _1 / |\varepsilon _1|)(\omega_1 + \delta_1)^* 
	\nonumber \\
	&& + i [ \cosh |\varepsilon _1| (\omega_2 + \delta_2)
	+ (e^{i \sigma_1} \varepsilon _1 / |\varepsilon _1|)
	(\omega_2 + \delta_2)^* ]]
	/ (\sinh^2 |\varepsilon _1|) ,
	\label{gamma1} 
\end{eqnarray}
and
\begin{eqnarray}
	\Gamma_2 &=& [i[ \cosh |\varepsilon _1| (\omega_1 + \delta_1)
	+ (\varepsilon _1 / |\varepsilon _1|)(\omega_1 + \delta_1)^* ]
	\nonumber \\
	&& + \cosh |\varepsilon _1| (\omega_2 + \delta_2)
	+ (e^{i \sigma_1} \varepsilon _1 / |\varepsilon _1|)
	(\omega_2 + \delta_2)^* ]
	/ (\sinh^2 |\varepsilon _1|) .
	\label{gamma2}
\end{eqnarray}
In the above calculation to obtain
eq.\ (\ref{squeezeintout4}), we have used the commutation relationship 
for $\hat{D}$ and $\hat{S}$ \cite{wallsmilburn}, as well as the 
relationship \cite{kimsanders}
\begin{eqnarray}
	&& \hat{B}_{ab} \hat{S}_a(\varepsilon ) \hat{S}_b(- \varepsilon )
	\hat{D}_a (\alpha ) \hat{D}_b (\beta ) \nonumber \\
	&& = \hat{S}_a (\varepsilon ) \hat{S}_b (- \varepsilon ) 
	\hat{D}_a([\alpha + i \beta ]/ \sqrt{2})
	\hat{D}_b ([\beta + i \alpha ]/ \sqrt{2}) \hat{B}_{ab} .
\end{eqnarray}

\section{Producing Entangled Coherent States with a Single Nonlinear Cell}

In Section V, it was demonstrated how entangled coherent states
could be produced with three nonlinear cells, without the need
for an interferometer.  One nonlinear cell
is used for the nonlinear interaction, and the other two 
are used to reverse-shear the state in each output.  However, a single
nonlinear cell, without the other two reverse-shearing cells, can be used 
by itself to create entangled coherent states, though the nature of
the output state has more complicated representation.

In the Fock state basis,
the output from a nonlinear cell with two coherent state inputs 
can be calculated using the nonlinear transformation in 
eq.\ (\ref{kerrop2}).  The result is 
\begin{equation}  
	\hat{\cal S} |\alpha \rangle _1 |\beta \rangle _2 
	\Bigr|_{\tau =0}
	= e^{-(|\alpha |^2 + |\beta |^2)/2}
	\sum_{m=0}^{\infty} 
	\sum_{n=0}^{\infty} 
	e ^{- i \chi m(m-1)} 
	e ^{- i \chi n(n-1)} 
	e^{-4i \chi mn} 
	\frac{\alpha ^m}{\sqrt{m!}}
	\frac{\beta ^n}{\sqrt{n!}}
	|m \rangle _1 |n \rangle _2 .
\label{twokerrin}
\end{equation}  
The output in eq.\ (\ref{twokerrin}) can also be expressed 
as an superposition of product coherent states.  This can be done
to obtain the result
\begin{equation} 
	\hat{\cal S} |\alpha \rangle _1 |\beta \rangle _2 
	\Bigr|_{\tau =0}
	= \int_0^{2 \pi } \frac{d \theta}{2 \pi }
	\int_0^{2 \pi } \frac{d \varphi}{2 \pi }
	g_{\chi } (\theta, \varphi )
	| \alpha e^{i (\chi - \theta )} \rangle _1
	| \beta  e^{i (\chi - \varphi )} \rangle _2 ,
\end{equation} 
where 
\begin{equation} 
	g_{\chi } (\theta, \varphi ) =
	\sum_{p,q=0}^{\infty} 
	\exp [-i (\chi p^2 - \theta p + \chi q^2 - \varphi q
	+ 4 \chi pq)] .
\end{equation} 

If $\chi / \pi $ is a rational number $2r/s$, then 
$|\alpha , \beta \rangle ^{\chi, \tau =0}$ can be expressed as a
finite sum of product coherent states,
\begin{equation} 
	\hat{\cal S} |\alpha \rangle _1 |\beta \rangle _2 
	\Bigr|_{\tau =0}
	= \sum_{m=1}^N \sum_{n=1}^N c_{mn} 
	|e^{i 2 \pi m/N} \alpha \rangle _1
        |e^{i 2 \pi n/N} \beta \rangle _2 .
\label{twokerrout}
\end{equation} 
As was true in eq.\ (\ref{entangledoutsimple}),
$N = s$ if $r$ and $s$ are relatively prime, and
$N < s$ is possible otherwise.
The coefficients $c_{mn}$ are found by solving the
simultaneous equations,
\begin{equation} 
	\sum_{m=1}^N \sum_{n=1}^N c_{mn}
	e^{i 2 \pi km/N} e^{i 2 \pi ln/N} 
	= e^{-i \chi [k(k-1) + l(l-1) + 4kl]},
\end{equation} 
for $k,l = 0,1, ..., N-1$.  This gives the result
\begin{equation} 
	c_{mn} = \frac{1}{N^2} \sum_{k=0}^{N-1} \sum_{l=0}^{N-1}
	\exp [-i 2 \pi (km + ln)/N - i \chi (k(k-1) + l(l-1) + 4kl) ] .
\label{twokerrcoeff}
\end{equation} 
If $2 \chi / \pi $ is an integer, the output will be a product state,
otherwise the output will be an entanglement of coherent states.

When $\chi = \pi /2$, we expect the output to be a product state.
Using eqs.\ (\ref{twokerrout}) and (\ref{twokerrcoeff}) 
yields the product state
\begin{equation} 
	\hat{\cal S} (\pi /2 , 0) \hat{D}_1 (\alpha ) \hat{D}_2 (\beta )
	|0 \rangle _1 |0 \rangle _2
	= - \frac{i}{2} 
	(|i \alpha \rangle _1 + i |-i \alpha \rangle _1)
	(|i \beta \rangle _2 + i |-i \beta \rangle _2).
\end{equation} 
For the case of a single nonlinear cell, the simplest entangled coherent 
state output is obtained for $\chi = \pi /4$:
\begin{eqnarray} 
	&& \hat{\cal S} (\pi /4 , 0) \hat{D}_1 (\alpha ) \hat{D}_2 (\beta )
	|0 \rangle _1 |0 \rangle _2 \nonumber \\
        && = \frac{1}{4} [ 
	i (|\alpha \rangle_1 - |-\alpha \rangle _1 )
	(|\beta \rangle _2 - |-\beta \rangle _2
	- e^{i \pi /4} |i \beta \rangle _2 
	- e^{i \pi /4} |-i \beta \rangle _2 ) \nonumber \\
        & & + e^{-i \pi /4} (|i \alpha \rangle_1 + |-i \alpha \rangle_1 )
	(|\beta \rangle_2 - |-\beta \rangle_2 
	+ e^{i \pi /4} |i \beta \rangle _2 
        + e^{i \pi /4} |- i \beta \rangle _2 ) ] .
\end{eqnarray}

\begin{figure}[t]
	\epsfysize=14cm
	\hspace{-2mm}
	\epsffile{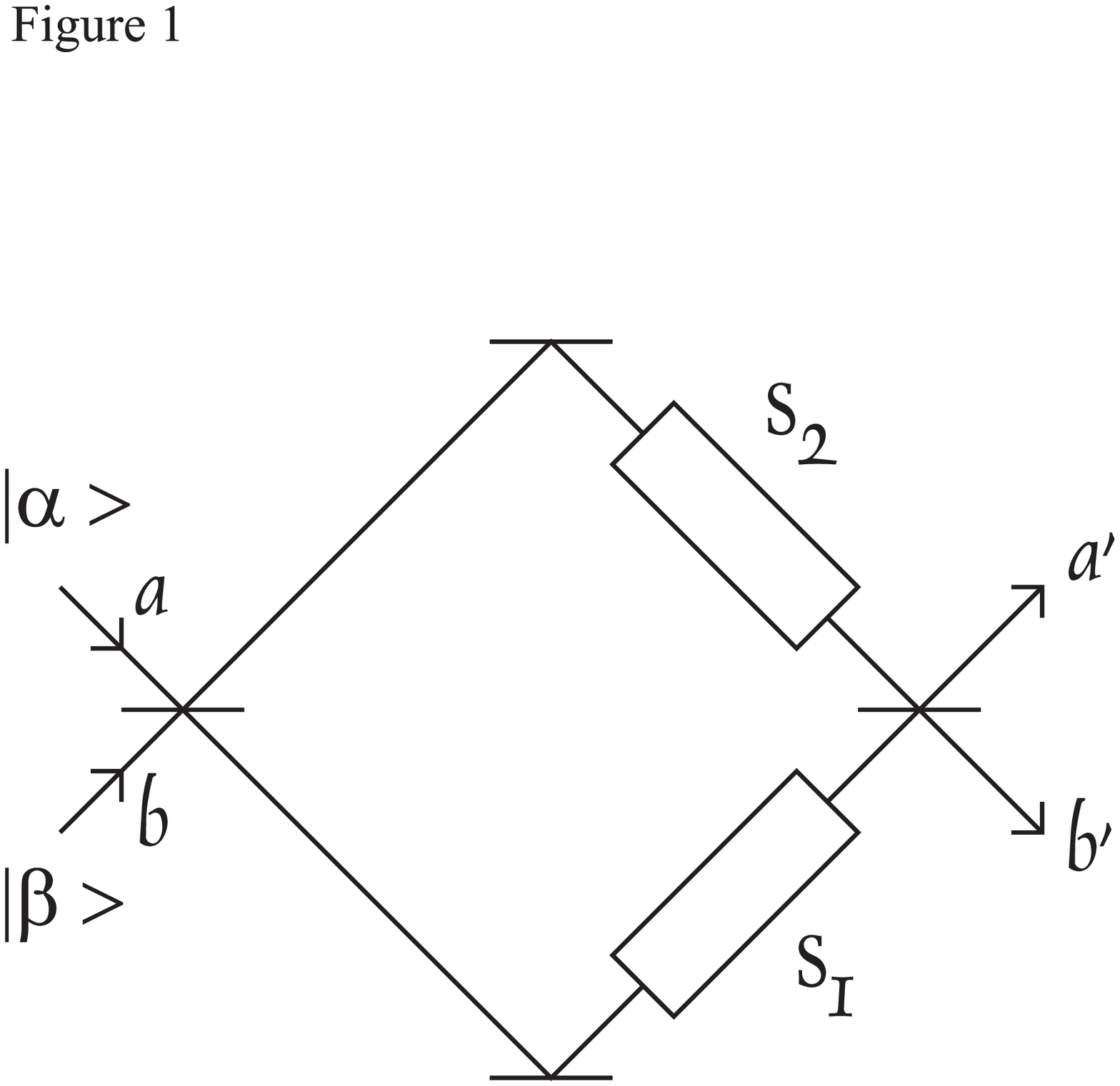}
	\vspace*{4cm}
	\caption{
	Nonlinear Mach--Zehnder interferometer.
	Coherent states $|\alpha\rangle$ and $|\beta\rangle$
	are injected into the two input ports of a beam splitter (BS).
	where they pass through a nonlinear medium.
	The fields are then recombined at the second BS.
	}
\end{figure}

\newpage

\begin{figure}[t]
	\epsfysize=16cm
	\hspace{-2mm}
	\epsffile{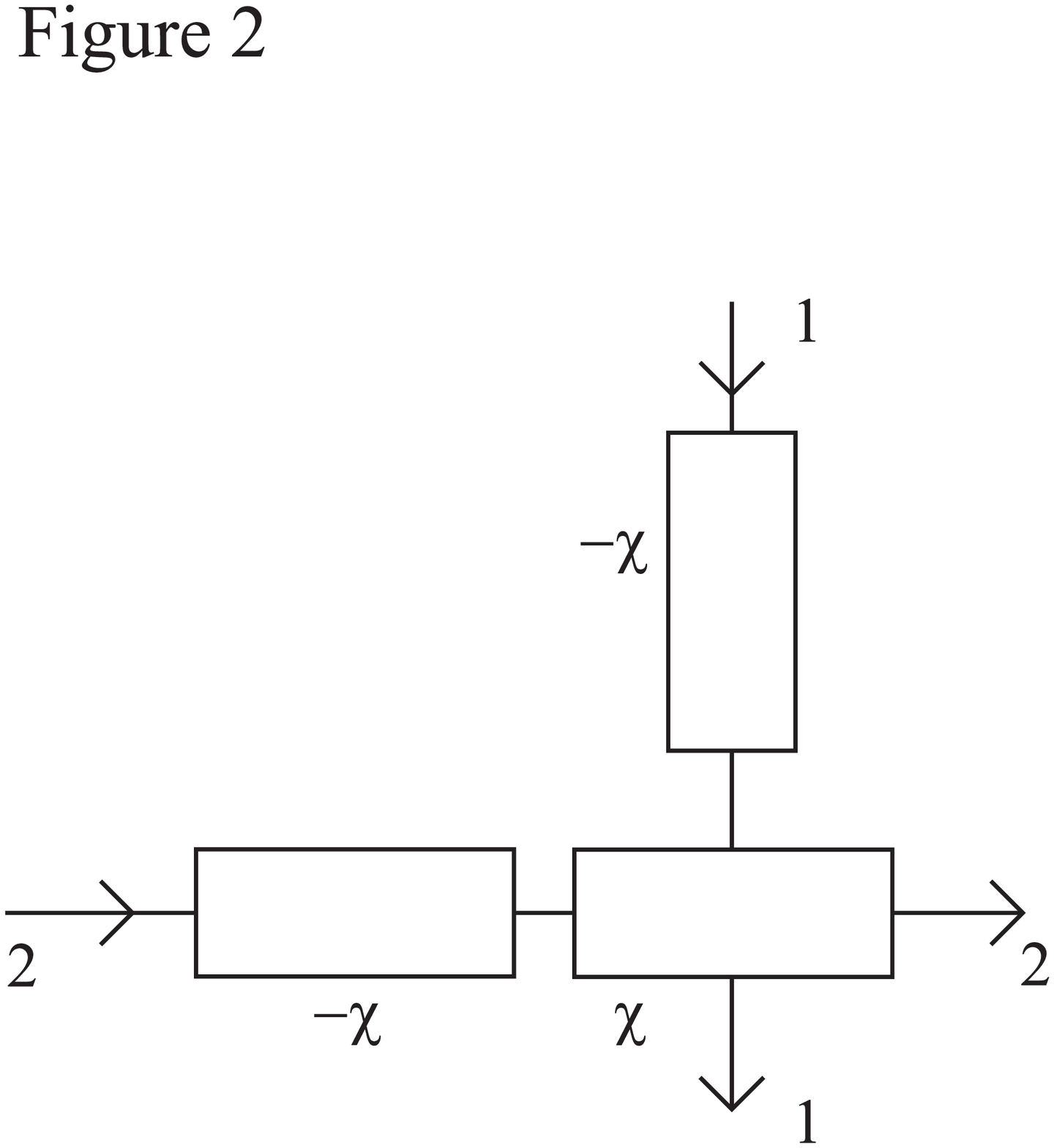}
	\vspace*{4cm}
	\caption{
	Three nonlinear media elements used to create an 
	entangled coherent state.
	}
\end{figure}

\end{document}